\documentstyle[psfig]{l-aa}

\def\arcsec{$^{\prime\prime}\,$}
\def\arcmin{$^{\prime}\,$}

\begin{document}

\title{Evidence for inverse Compton emission in the 
powerful radio galaxy 3C 219}

\thesaurus{
03(02.18.5; 11.01.2; 11.09.1 3C 219; 11.13.2; 13.18.1; 13.25.2)}

\author{G. Brunetti\inst{1,2} 
\and A. Comastri\inst{3} 
\and G. Setti\inst{1,2}
\and L. Feretti\inst{2} }
\institute{Dipartimento di Astronomia, 
via Zamboni 33, I--40126 Bologna, Italy
\and
Istituto di Radioastronomia del CNR,
via Gobetti 101, I--40129 Bologna, Italy 
\and
Osservatorio Astronomico di Bologna,
via Zamboni 33, I--40126 Bologna, Italy}


\offprints{G.Brunetti, Istituto di Radioastronomia del CNR, via Gobetti
101, I--40129 Bologna, Italy}

\maketitle
\markboth{G. Brunetti et al.}{Inverse Compton emission from 3C 219}

\begin{abstract}
Spectral analysis of ROSAT PSPC and ASCA archive data and a recent ROSAT HRI
observation of the powerful FRII radio galaxy 3C 219 reveal an absorbed
point--like source, coincident with the nucleus, and a non--thermal
extended component aligned with the radio structure.
The point--like source can be readily interpreted as a hidden quasar in
the nucleus, giving further support to the unification scheme of
FRII radio galaxies and radio loud quasars.
The coincidence between  the X--ray (0.1--10 keV) and radio spectral
slopes suggest that most of the emission associated with the extended component
is due to the inverse Compton process in the radio lobes.
The extended circumnuclear emission can be understood as Compton scattering
of the IR--optical radiation emitted by the hidden quasar and surrounding
dusty/molecular torus.
This is the first observational evidence supporting the existence of this
effect, which also
probes the relativistic electron spectrum at energies
much lower than those involved in the synchrotron radio emission.
The observed X--ray flux can be matched by assuming that the energy density
of the relativistic particles exceed the equipartition value by about
a factor of 10.
At larger distances from the nucleus ($\geq 70$kpc) the inverse Compton
scattering
with the cosmic microwave background photons becomes more important
and may explain the observed X--ray features if positive
fluctuations in the column densities of relativistic electrons
are present.
Alternatively, one cannot exclude with the present data a thermal contribution
by hot clumpy gas surrounding the radio lobes.
\end{abstract}



\section{Introduction}

\par

The properties of the X-ray emission of the
powerful radio galaxies have been
investigated in the literature
(Fabbiano et al. 1984, Crawford \& Fabian 1995, 1996).

\noindent
In the framework of the AGNs unification models (Barthel 1989,
Urry \& Padovani 1995) the nuclear X--ray emission of the 
FRII radio galaxies is expected to be heavily absorbed 
along the line of sight by a circumnuclear molecular
torus.
Strong absorption has indeed been detected in the  
X--ray observations of the narrow line FRII radio galaxies Cyg A 
(Ueno et al. 1994) and 3C 194 (Crawford \& Fabian 1996); moreover, 
intrinsic absorption has been discovered 
in the X--ray spectra of the broad line radio galaxies (BLRGs)
3C 109 (Allen \& Fabian 1992) and  3C 287.2 (Crawford \& Fabian 
1995).

\noindent
Worrall et
al. (1994) have suggested that part of the soft X--ray emission
of some strong FRII radio galaxies is
synchrotron self--Compton (SSC) in the AGN jets.
SSC may also explain the X--ray emission observed with ROSAT in 
two emission regions coincident with the radio hot--spots 
of Cygnus A (Harris et al. 1994), but it is generally
too weak to be detected even in the nearby radio galaxies 
(Hardcastle et al. 1998a).

\noindent
It is well known that
non--thermal mechanisms can produce extended X--ray emission in 
the lobes of the radio galaxies where the relativistic electrons can
interact with the microwave background photons (CMB) and radiate
via the inverse Compton (IC) process.
The relativistic particle densities and magnetic field strengths in the 
extended lobes of radio galaxies are usually estimated on the basis of the
minimum energy assumption (equipartition).
The detection of IC 
scattered X--rays from the radio lobes could
provide an invaluable tool to determine the value of these important physical
parameters. Unfortunately, this process is not particularly efficient at low
redshifts (the IC emissivity increases as $\sim$(1 + z)$^4$) so that ROSAT and
ASCA have failed to detect this effect in nearby radio galaxies with a
few exceptions (notably, Fornax A: Feigelson et al. 1995, Kaneda et al. 1995;
Cen B: Tashiro et al. 1998).

\noindent
In a recent paper Brunetti et al. (1997) have shown that 
in the
framework of the unification scheme linking
strong FRII radio galaxies 
and radio loud quasars it is possible to predict large X--ray
fluxes by the IC scattering of the far/near IR photons from a typical
``hidden'' quasar with the relativistic electrons in the radio lobes. 
Even if this effect is expected to dominate the IC contribution
also at high redshifts, and 
predicts X--ray fluxes and spectral shapes
consistent with those observed for a number of strong
distant FRII radio galaxies (redshift z $\sim$ 1), it is
difficult to be tested in detail because of the low 
spatial resolution and sensitivity
of the available X--ray telescopes.
The far/near IR photons from the ``hidden'' quasar are scattered in the X--ray 
band mainly by mildly
relativistic electrons ($\gamma\sim 100-300$) which are not those responsible
for the synchrotron radio emission (typically $\gamma \geq 5000$).
Therefore, the detection of the IC emission predicted by Brunetti et al.(1997)
model
would not only provide additional evidence in favour of the 
radio galaxy/quasar unification scheme, but it would also provide an
unique handle to probe the spectral and spatial distributions of the 
relativistic particles over a wider energy range.

Nevertheless, optical and radio observations have suggested that 
relatively distant radio galaxies (z$\geq$ 0.3) lie at the center
of moderately rich clusters 
of galaxies (Yates et al. 1989, Hill \& Lilly 1991)
with a dense hot intracluster medium (Garrington \& Conway 1991).
In addition,  X--ray data are consistent with a scenario
in which more than two third of the 50, or so, nearby brightest clusters  
have cooling flows (Fabian 1994).
As a consequence in a number of powerful radio galaxies 
the thermal emission from the intracluster medium 
is expected to dominate the observed soft X--ray flux,
as in Cyg A (Reynolds \& Fabian 1995), producing extended
X--ray structures and shadowing any other contribution.
This scenario is further complicated by the fact that
deep  ROSAT HRI observations of nearby radio galaxies 
have revealed X--ray deficit associated with the radio lobes possibly
originating from the hydrodynamical
interaction of the jets and radio hot--spots with the intracluster medium
(Cyg.A: Carilli et al. 1994; NGC 1275: B\"ohringer et al. 1993; 3C 449:
Hardcastle et al. 1998b).

We present here a deep ROSAT HRI observation of 3C 219, a 
nearby (z=0.1744) powerful radio source identified
with a cD galaxy 
of magnitude M$_V$=-21.4 (Taylor et al.1996), belonging to a non Abell cluster
(Burbidge \& Crowne 1979).
Despite of its cluster membership, the
ROSAT PSPC and ASCA archive data suggest that thermal emission, if
present, is negligible so that this source could be a good candidate to 
detect possible IC contribution to the X--ray flux.

The radio structure is well studied 
(Perley et al. 1980,
Bridle et al. 1986, Clarke et al. 1992): it is a classical double--lobed
FRII radio galaxy 
that spreads over $\sim$ 180\arcsec on the sky plane corresponding to a
projected size $\sim$ 460 Kpc.  
\footnote{{$H_0=75$km s$^{-1}$Mpc$^{-1}$ and $q_0=0.0$ are assumed
throughout}}
A strong jet extends over 20\arcsec 
($\sim$ 50 Kpc) through the south--western lobe, while a weak counterjet is
visible in the north--eastern lobe.
The total radio spectrum,
largely dominated by the extended radio structure,
between 178 and 750 MHz is $\alpha=0.81$
($S_{\nu}\propto \nu^{-\alpha}$, Laing et al. 1983).

Fabbiano et al. (1986) have reported the detection of a broad
Paschen--$\alpha$ line in excess of the predictions of case B recombination
implying the presence of an absorbed broad line region with 
$A_V>1.6$ mag.
More recent optical studies have shown that the 3C 219 spectrum is dominated
by a starlight continuum, but a non--stellar component and emission line
features are also present (Lawrence et al. 1996);
from the broad line ratio $H_{\alpha}/H_{\beta}$,
assuming typical broad line region parameters, we infer
$A_V\sim2.5$.
In a H$\alpha$+[N II] image taken with the 4m Kitt Peak telescope
the galaxy appears to be point--like 
(Baum et al. 1988) with a nearby, possibly interacting companion 
$\sim 10^{\prime \prime}$ to the south--east.
An image of 3C 219 has been 
taken with the HST telescope's WFPC2 through a broadband red (F702W)
filter (de Koff et al. 1996); the galaxy is well resolved and does not
present relevant features (dusty lane, distortions) on a scale 
$\geq 300$ pc (i.e. HST resolution).

The data analysis is presented in Sect. 2 and 3
while the proposed interpretation is
discussed in Sect.4.

\section{X--ray spectral analysis}

\subsection{ROSAT PSPC observations}

\par

\begin{table*}
\caption{Spectral fits with ROSAT and ASCA data (f indicates a frozen 
parameter)}
\begin{tabular}{lllllllll}
\hline 
\noalign{\smallskip}
Instrument & En. Range & N$_H^a$ & $\Gamma^b$ & N$_H^c$ & $C_V^d$ 
& $E_{line}^e$ & $EW^g$ & $\chi^{2}/dof^h$ \\
\noalign{\smallskip}
\hline
\noalign{\smallskip}
ROSAT & 0.1--2.4 keV & $1.8\pm1.1$ & $1.17\pm0.37$ & --  & -- & -- & -- &
7.6/10 \\
    &       &  1.55(f) & $1.09\pm0.14$ & -- & -- & -- & -- & 7.7/11 \\
    &     &   1.55(f) & 1.80 (f) & 18.7$_{-13.1}^{+19.6}$ & 0.74$\pm$0.07 
    &  -- & -- & 9.9/10 \\
\noalign{\smallskip}
ASCA & 0.6--10 keV & $22.6_{-5.3}^{+5.7}$ & $1.78\pm0.08$ & -- & -- 
& -- & -- & 298.5/283 \\
     &   & 1.55(f) & $1.51\pm0.04$ & -- & -- & -- & -- & 359.2/284 \\
    &    &   $23.6_{-5.4}^{+5.9}$ & $1.81\pm0.08$ &
    -- & -- & 6.4(f) & $<$ 224 & 295.8/285 \\
\noalign{\smallskip}
ROSAT+ASCA & 0.1--10 keV & 1.55(f) & $1.75\pm0.08$ & $21\pm 7$ 
& 0.74$\pm$0.08 & -- & -- & 304.8/293 \\
\noalign{\smallskip}
\hline
\noalign{\smallskip}
\end{tabular}

$^a$ Equivalent hydrogen column density (units of $10^{20}$ cm$^{-2}$) \\
$^b$ Photon spectral index \\
$^c$ Column density of the partial covering gas ($10^{20}$ cm$^{-2}$) \\
$^d$ Covering Fraction \\
$^e$ Iron line energy (keV) \\
$^g$ Iron line EW (eV) \\
$^h$ Total $\chi^2$ and degrees of freedom 
\end{table*}
3C 219 has been observed with the ROSAT PSPC (Pfeffermann et al. 1986)
on May 3, 1992.

We have analyzed the archive data finding that
the spatial profile in the 0.1--2.4 keV range is consistent 
with that of a point source convolved with the PSPC point spread function 
(PSF), taking into account
the source spectral properties and  
the background level. However, the maximum likelihood detection algorithm 
suggests some possible evidence of extended emission (see Sect.3).
There is no evidence of significant time variability.

The source spectrum was extracted from a circular region 
with 2\arcmin radius.
Different background spectra have been extracted
either from annuli centered on the source
or from circular regions uncontaminated by nearby sources.
In all cases the background appears stable without 
any appreciable variations within the statistical errors.
Corrections were included for vignetting and PSPC dead time.
Source spectra were extracted 
in the pulse invariant (PI) channels in the range 11--240 
($\simeq$ 0.1--2.4 keV). 
The photon event files were analysed using the EXSAS/MIDAS 
software (version 94NOV, Zimmermann et al. 1993) and the extracted 
spectra were analysed using version 9.0 of XSPEC (Shafer et al. 1991)
with the appropriate response matrix.
The resulting exposure time is 4349 s and the
background subtracted source count rate is 0.127 $\pm$ 0.006 cts s$^{-1}$.
The source spectrum was rebinned in order to obtain a significant
signal to noise ratio ($>5$) for each bin and fitted either with a power law 
or with a thermal plasma spectrum 
(Raymond--Smith model) plus
absorption arising from cold material with solar
abundance (Morrison \& McCammon 1983).
The derived spectral parameters are given in Table 1, where the reported
errors are at 90\% confidence level
(Lampton et al. 1976);
the value of the Galactic column
density towards 3C 219, $N_H =1.55\times 10^{20}$ cm$^{-2}$,
has been retrieved from the 21 cm radio survey 
of Dickey \& Lockman (1990).
A single power law with the absorption either fixed at 
the Galactic value, or free to vary, provides an acceptable 
description of the observed  spectrum (Table 1).
Formally, also a Raymond--Smith thermal model with solar abundance
fits well the observed counts, however the derived temperature
is extremely high and almost unconstrained.

The best fit spectral photon index is unusually flat
in agreement with the results of Prieto (1996).
The flat power law slope can be due to the effect of absorption 
on a steeper continuum. In order to test this possibility we have tried
a partial covering model. 
Assuming a typical AGN spectrum with photon index $\Gamma$=1.8 
plus Galactic absorption, a good description of the data 
can be obtained if about 75\% of the nuclear radiation is absorbed
by a column density of the order of $\sim$ 2 $\times 10^{21}$ cm$^{-2}$ 
(Table 1).
The X--ray flux in the 0.1--2.4 keV energy range, corrected for
Galactic absorption, is 
1.8 $\times$ 10$^{-12}$ ergs cm$^{-2}$ s$^{-1}$ corresponding to an
isotropic  
luminosity of 1.1 $\times$ 10$^{44}$ ergs s$^{-1}$ (rest frame),
while the unabsorbed nuclear luminosity 
would be $3.6\cdot 10^{44}$erg s$^{-1}$.

\subsection {ASCA Observations} 

\par

3C 219  was observed with ASCA 
(Tanaka, et al. 1994) 
on April 12, 1995 and on November 13, 1994 
with the Gas Imaging Spectrometers (GIS2/GIS3) 
and with the Solid--state Imaging Spectrometers (SIS0/SIS1).
Both observations were analyzed by us using standard calibration and
data reduction methods (FTOOLS) provided by the ASCA Guest Observer 
Facility at Goddard Space Flight Center.
The net exposure time for the 1995 observation was 18 Ks in the GIS detectors
and 16.5 Ks in the SIS. Slightly lower exposure times were obtained for the
1994 observation.

\begin{figure*}
\centerline{
\psfig{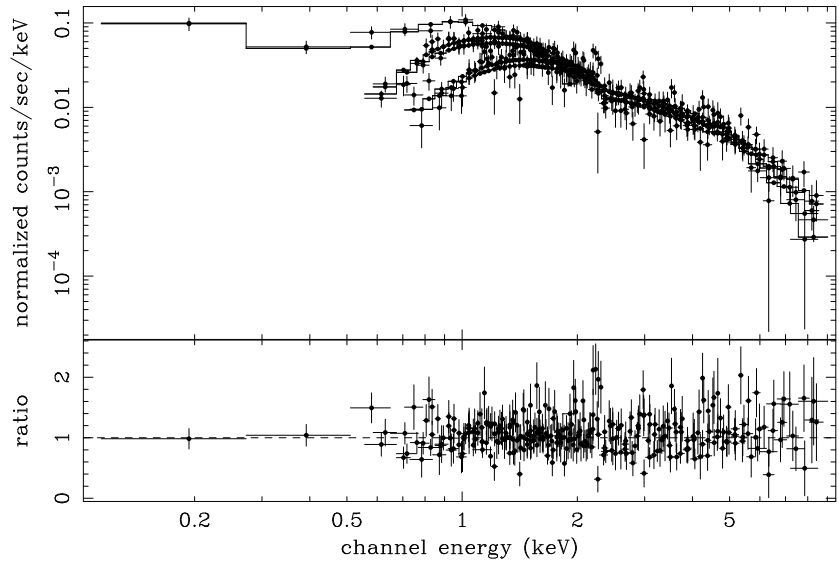}
}
\caption{
3C 219 spectrum from 
ROSAT PSPC, ASCA GIS and SIS joint fit.
The fit has been obtained by leaving the relative normalizations free to
vary (see text).
The best fit is a partial covering model with $\Gamma=1.75\pm 0.08$,
covering fraction$=0.74\pm 0.08$, 
and $N_H=2.1\pm 0.7 \times 10^{21}$cm$^{-2}$.
This model accounts for both the high energy (ASCA) part and 
the more flat low energy (ROSAT) part of the X--ray spectrum.
The normalized cts s$^{-1}$ keV$^{-1}$ 
are shown in the upper panel, while the
ratio between the observed points and the model
is shown in the lower panel; no iron line is requested
by the fit.}
\end{figure*}

Source photons were 
extracted from a circular region centered on the source with 6$^{\prime}$
radius for GIS and 3$^{\prime}$ radius for SIS.
The background was estimated using both source-free regions 
and blank--sky GIS and SIS observations available in the
calibration area. Different background estimates give
consistent results. 

Data preparation and spectral analysis were performed using version 1.3
of the XSELECT package and version 9.0 of the XSPEC program.
The light curves from each instrument do not show any 
significant flux variation over the whole observation.
GIS and SIS spectra were binned with more than 20 cts/bin
in the 0.7--10 keV and 0.6--10 keV energy ranges respectively.
The lowest SIS energy channels have been excluded because of the
uncertain calibrations (Dotani et al. 1996; Cappi et al. 1997).
Since the spectral parameters obtained by fitting the four detectors
separately were all consistent within the errors, data 
from both pair of SIS and GIS 
were fitted simultaneously to the same model, but with the normalization
of each dataset allowed to vary relative to the others 
in order to account for the small discrepancies in the 
absolute flux calibrations of the detectors.

Both observations gave very similar results; in the following we
discuss only the results obtained from the 1995 observation characterized by
a better counting statistics.

A single power law model clearly provides an acceptable fit 
to the data (Table 1). Absorption in excess of the 
Galactic value is required.
There is no need of more complex models. In particular the addition of 
a narrow emission line at 6.4 keV does not improve the quality
of the fit, while
a thermal component is not required by the data.
The absorption corrected X--ray flux in the 2--10 keV
energy band is $\sim$ 3.6 $\times 10^{-12}$ erg cm$^{-2}$ s$^{-1}$, 
which corresponds to a luminosity of  $\sim$ 2.4 $\times 10^{44}$ erg s$^{-1}$
in the source frame.

\subsection{ROSAT and ASCA joint fits}

\par

In the overlapping 0.6--2.0 keV energy range the observed PSPC flux
is about 15 \% lower 
than the ASCA flux with a weak dependence on the assumed spectral 
parameters.
Moreover the value of the column density derived from the PSPC data 
assuming a partial covering model (Table 1) is in good agreement with 
the absorption observed in the ASCA data (Table 1) suggesting that 
this model is viable or that two different spectral components
are present.
Therefore, in order to make full use of the available spectral band 
(0.1--10 keV) we have performed joint fits to the ROSAT PSPC and ASCA data
leaving the relative normalizations free to vary to account for the 
residual absolute flux uncertainties among the different instruments.

Not surprisingly a partial covering model provides a good description
of the observed 0.1--10 keV spectrum:
the resulting power law spectrum
($\Gamma=1.75 \pm 0.08$) is partially ($74 \pm 8$ \%) absorbed by cold 
gas with a column density $N_H = 2.1 \pm 0.7 \times 10^{21}$ cm$^{-2}$
(Fig.1, Table 1).
This implies that in the soft 0.1--2.4 keV band about 38\% of the
observed flux may be due to a scattered component.
A description of the soft X--ray spectrum in terms of thermal 
emission is not viable, while the addition of a thermal 
component to the best fit partial covering model does not 
significantly improve the fit.
Any thermal component (if present) cannot contribute more than 8--10 \% 
(90 \% confidence limit) to the observed X--ray flux.

We conclude that
the 0.1--10 keV spectrum of the radio galaxy 3C 219 can be explained 
in terms of an obscured central source 
characterized by a power law
with a slope typical of radio loud quasars in the 2--10 keV 
energy range (Lawson \& Turner 1997). 
An additional unabsorbed spectral component 
is present in the soft X--ray band. 
In principle
this component may be due to
scattering of the soft X--ray nuclear radiation
by circumnuclear clouds or thermal electrons,
but we note that the unabsorbed power law 
(Table 1)
has a slope close to that of the
synchrotron radio spectrum ($\alpha=0.81$),
so that the possibility
that it might originate from IC scattering of the IR--optical nuclear 
and/or CMB radiation can also be taken into account.
A crucial test to decide which one 
of these two possibilities should be adopted 
is the study of the spatial distribution of the X--ray flux.

\section{Spatial Analysis}

\par

3C 219 was observed with the ROSAT HRI for a total of 
28.6 Ks between April 10 and May 8, 1997.
Standard procedures have been employed within the MIDAS/EXSAS software.
The source counts, estimated with a maximum likelyhood (ML) detection algorithm 
as described in Cruddace et al. (1988), are 847$\pm$30 
corresponding to a background subtracted count rate of $\sim$ 0.03 
cts s$^{-1}$ in the 0.1--2.4 keV band.
Since the ML algorithm has been conceived to study  point--like
sources and is not efficient for extended sources, the source flux  
has been computed measuring the counts in a circle of 2\arcmin radius.
We find evidence of emission
outside 20\arcsec from the nucleus
accounting for some 15\% of the total flux (see Sect.4).
The resulting  
total count rate (nuclear and extended emission) is $\sim$ 0.035 cts s$^{-1}$. 
The total HRI flux of $\sim$ 1.6 $\times 10^{-12}$ erg cm$^{-2}$ s$^{-1}$
is consistent, within about 10\%, with 
that derived from the PSPC observation (see $\S$ 2.1) assuming  
the spectral parameters reported in Table 1.

It is known that the intrinsic spatial resolution of the ROSAT
HRI of $\sim$5\arcsec FWHM is blurred by the errors due to the relatively poor
knowledge of the pointing position as a function of time.
We have tried to correct for the residual aspect solution errors
following a procedure developed by Harris et al.(1998)
which selects observing periods with the same roll angle
and folds the data according to the wobbling period of 402 s.
Several subimages are created according to the source statistics and the
wobble period, then shifted to a common center and coadded.

\begin{figure}
\centerline{
\psfig{figure=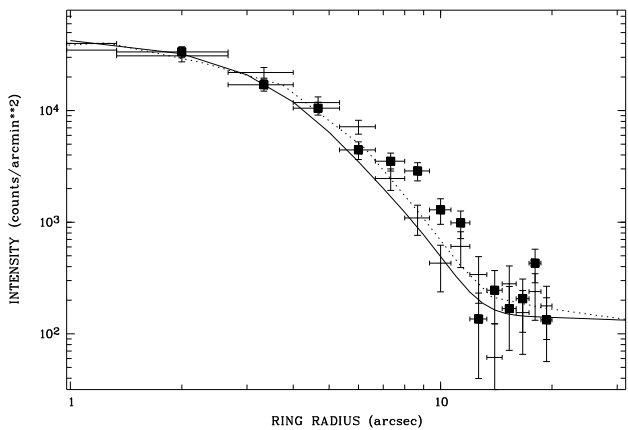,width=9cm,angle=270}
}
\caption{
The wobble corrected ROSAT HRI radial profile in the
0.1--2.4 keV band.
The counts have been taken from two symmetric sectors: one
in the north--south direction (filled squares) from 
-45$^{\circ}$ to 45$^{\circ}$ and from 135$^{\circ}$ to 225$^{\circ}$
clockwise starting from the north, and
one in the east--west direction from
45$^{\circ}$ to 135$^{\circ}$ and from 225$^{\circ}$ to 315$^{\circ}$.
The expected profiles of a point source according to the theoretical
PSF (continuous line) and to the inflight calibrated PSF (dotted line)
are reported normalized to the source intensity and background level.}
\end{figure}

The radial profile of the innermost region
of the wobble corrected image is then compared 
with an inflight calibrated PSF,  
kindly provided by I. Lehmann (private communication), derived
from the average profile of 21 bright stars (Fig.2).
From the analysis of the 
azimuthal distribution of the counts within
$\sim$20\arcsec ($\sim$ 50 kpc) from the source peak intensity we find
that the nuclear source is resolved in the
north--south direction (Fig.2).
A Kolmogorov--Smirnov test shows
that the observed distribution of the counts 
in the north--south direction 
differs from the PSF 
at the 99\% confidence level, while the test is not conclusive
in the case of the east--west direction.

In order to get more information on the structure an X--ray image
of 3C 219, smoothed with a circular gaussian of
FWHM=3\arcsec (Fig.3), has been produced with the EXSAS package, while 
the further image processing has been performed with the Astronomical
Image Processing System (AIPS) package.
An extended structure
up to $\sim$15\arcsec in radius is indeed detected 
in the north--south direction (Fig.3).

\begin{figure*}
\centerline{
\psfig{figure=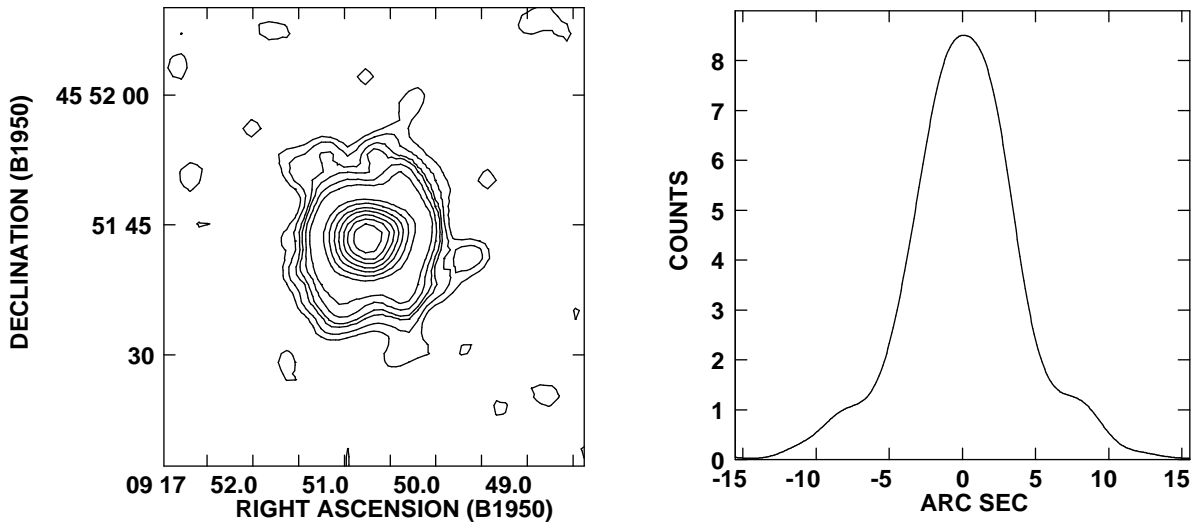,width=10cm}
}
\caption{
The X-ray image of the bright central source coincident with 3C 219,
smoothed with a Gaussian of FWHM = 3\arcsec, is shown in the left panel. 
Contour levels are 0.15, 0.25, 0.4, 0.5, 0.7, 1, 2.6, 3.4, 4.3, 5.1, 
6.8, 7.7 cts pixel$^{-1}$, with the peak corresponding to 8.56 
cts pixel$^{-1}$ (1 pixel = 1\arcsec$\times$1\arcsec).
The asymmetry of the central source (highest contour levels) can be
attributed to a slight degradation of the PSF
(FWHM=8.2\arcsec$\times$7.5\arcsec with a position angle
of $\sim$70$^{\circ}$ clockwise from N), while the
lower level structure is likely to be due to real source extension.
This is confirmed by the profile across the source (right panel), obtained
at the position angle of 28$^{\circ}$ (clockwise from N),  which clearly
shows that the excess emission at distance of 7--12\arcsec
from the peak cannot be due to the PSF.}
\end{figure*}

The smoothed PSF appears to be elliptical 
(FWHM= 8.2\arcsec$\times$7.5\arcsec) with the major axis positioned at
$\sim$70$^{\circ}$ clockwise from the north, at 
variance with the elongation of the 
extended structure.

Of course, the small FWHM used in the previous smoothing procedure does not
allow a detection of the outer extended
low brightness features.
The X-ray image of Fig.4
has been produced with the EXSAS package 
by binning the photon event table in
pixels of 2\arcsec and by smoothing the map with a circular gaussian
of $\sigma$ = 5\arcsec. With these values the signal to noise ratio
is good enough to image the low brightness extended emission
significantly above the background.
The ROSAT HRI image has been shifted by 1.5\arcsec~ and
0.6\arcsec~ in right ascension and declination, respectively, in order
to align the X-ray peak with the radio core position.
The X--ray map has been overlaid to the 
optical POSS--II digitized image of the field in Fig.4.
Since 3C 219 is in a cluster of galaxies,
the optical image is crowded and a number of objects fall within the
X--ray structure; no coincidence with relevant optical
objects is found.

\begin{figure}
\centerline{
\psfig{figure=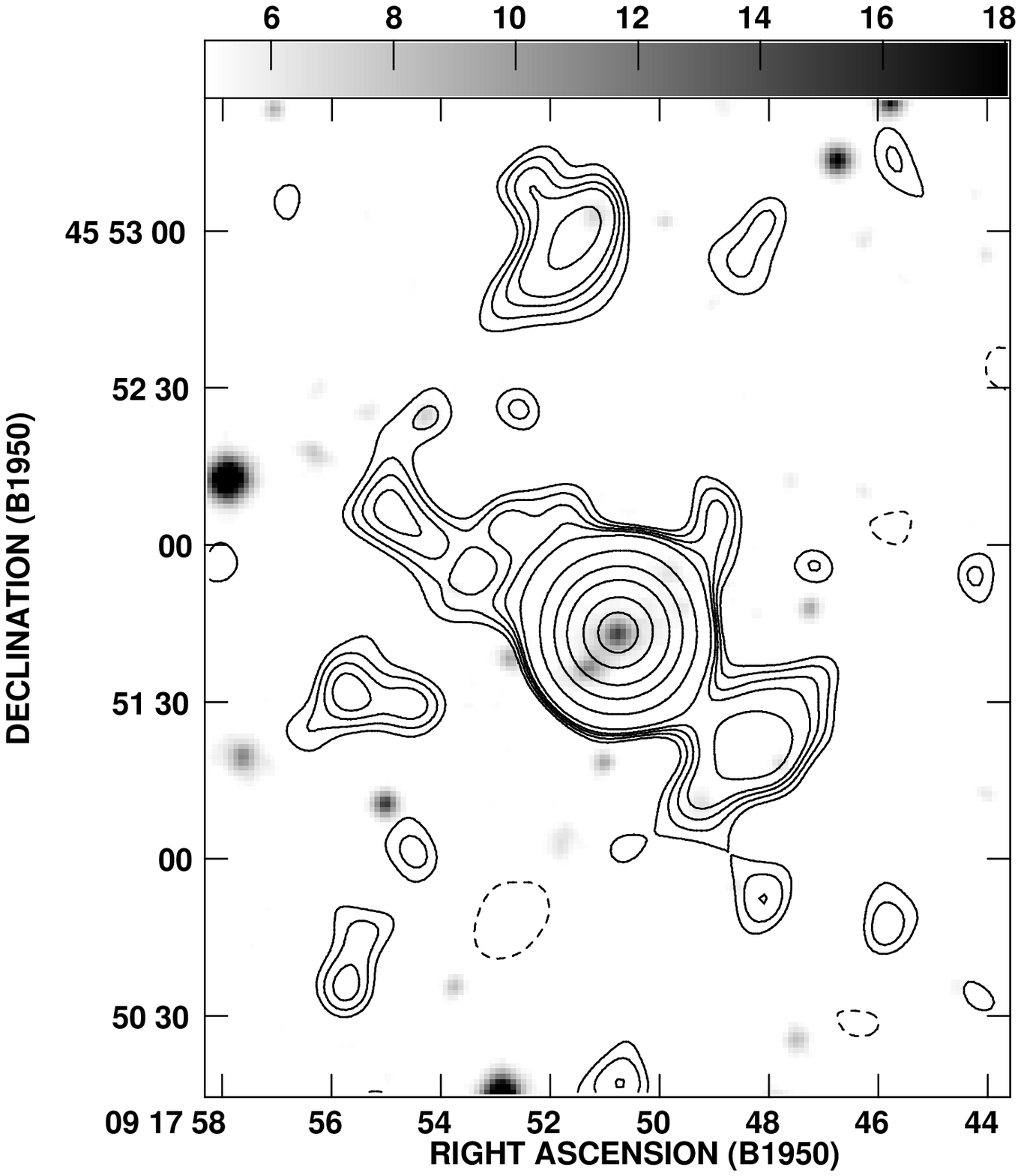,width=9cm}
}
\caption{
The X--ray image of 3C 219 (contours)
superposed on the red optical POSS-II digitized
plate (gray--scale).
The X--ray flux is
dominated in the central region by a point--like source, 
but an extended component is detected up to $\sim$90 \arcsec from the
nucleus.
The X--ray image has been produced by binning the photons 
in pixels
of 2\arcsec and by smoothing the map with a circular gaussian of 
$\sigma$= 5 \arcsec.
The plotted contour levels are: 0.05 (dashed), 0.2, 0.22, 0.24, 0.26,
0.31, 0.65, 1.5, 3, 6, 9 cts pixel$^{-1}$ (1 pixel = 
2\arcsec$\times$2\arcsec), with the first two contours 
respectively at the 2--$\sigma$ level below 
(dashed) and above the background.
The background is 0.125 cts pixel$^{-1}$. 
The optical image also shows 
a companion galaxy at $\sim$ 10\arcsec S--E of 3C 219.}
\end{figure}

The structure of the X-ray brightness distribution between
$\sim$15--25\arcsec~ from the core can be enhanced if the central
unresolved component is subtracted from the HRI image.  The
model used in the subtraction is a circular gaussian ($\sigma$ =
6\arcsec) with the peak coincident with the X-ray peak.  
This function represents the convolution between the instrument PSF and the
smoothing function applied to the image.  In fact, although the instrument 
PSF is not exactly a gaussian, the smoothing gaussian dominates the
resulting convolved PSF.  

In order to subtract the nuclear component from the total HRI image
we have made use of the spectral analysis results of Sect.2.
The 0.1--2.4 keV flux of the nuclear emission
has been estimated by assuming the best fit spectral parameters 
of the high energy absorbed component in the 0.1--10 keV fits.
Taking into account 
the different responses 
of the HRI, PSPC and ASCA 
detectors for a given spectrum, as well as the errors 
on the best fit spectral parameters, 
the nuclear flux is estimated to account from 55\% to 70\%
of the HRI counts.
The subtraction procedure has been performed several times 
constraining the amplitude of the nuclear component 
within this interval.
We also note that
the largest contribution of the point--like nuclear source
(leaving zero counts after
the subtraction at the position of the peak)
is $\sim 74$\% of the total net
counts; this limit is implied from the HRI data alone.
The residual map resulting from the subtraction of the nuclear source
with a representative amplitude of 64\% is shown in Fig.5 superposed on the 
VLA radio image at 1.4 GHz (Clarke et al. 1992).
\begin{figure}
\centerline{
\psfig{figure=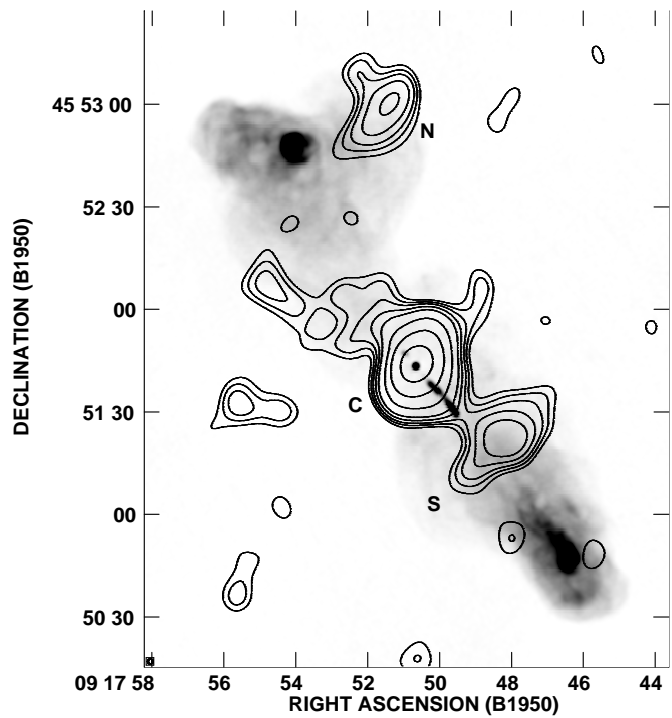,width=9cm}
}
\caption{
The X--ray image of 3C 219 (contours) after subtraction
of the central point--like
source superposed on the VLA radio image at 1.4 GHz with 1.4\arcsec
resolution (gray--scale).
The contours are:
0.22, 0.24, 0.26, 0.29, 0.33, 0.45, 0.65, 1 cts pixel$^{-1}$
(1 pixel= 2\arcsec$\times$2\arcsec).
The conversion from counts to X--ray brightness 
(erg cm$^{-2}$ s$^{-1}$ arcsec$^{-2}$) is
$4.1\cdot 10^{-16}$ for a power law 
spectrum ($\Gamma =1.8$) or  $3.4\cdot 10^{-16}$ for a
thermal spectrum ($kT =1.5$ keV).
The C--component contains 
about 60\% of the counts.}
\end{figure}
It is evident that
the residual X--ray isophotes
are strongly correlated with the radio extension and appear
to be elongated along the radio structure.
We distinguish three main regions: one coincident with the 
low brightness part of the northern radio lobe (N), one positioned between
the nucleus and the southern radio hot--spot (S) and the strongest one (C) 
centered on the radio--core.
The C--component appears like a curious eight--shaped figure 
with the axis inclined by $\sim 40^{\circ}$ with respect
to radio--axis, but the isophotes become more aligned with the 
radio--axis with increasing distance from the nucleus.
This basic
structure persists even by varying the amplitude of the 
point--like source within the allowed interval and/or
by applying a slightly different $\sigma$ in the subtraction
procedure.
Furthermore one can notice that the C--component seems to show a remarkable
continuity with the extended structure of Fig.3.

\section{The Model}

\par

The ROSAT HRI observation presented here shows a complex
extended X--ray structure surrounding a central point--like
source.
ROSAT and ASCA spectral analysis reveals that the nucleus is
absorbed by a column density 
$N_H \sim 2-3\cdot 10^{21}$cm$^{-2}$ similarly  
to other ROSAT findings for BLRGs
(Crawford \& Fabian 1995).
The unabsorbed luminosity 
($\simeq 3.6\cdot 10^{44}$erg s$^{-1}$)
in the 0.1--2.4 keV band and the photon index 
$\Gamma \simeq 1.7-1.8$ are typical of steep spectrum radio loud quasars, 
thus strengthening the hypothesis that we are dealing with a hidden
quasar in the nucleus of the radio galaxy in agreement with the
unification model.

According to the spectral analysis,
the nuclear source contributes $\sim 62$\% of the total
ROSAT PSPC flux while the contribution from a thermal source, if any, cannot
exceed $\sim 10$\% of the total flux.
Therefore a large fraction, if not all,
of the flux associated with the extended structure
should be characterized by a power law spectrum with a slope
close to that of the radio emission ($\alpha\simeq 0.8$).
This and the spatial correlation with the overall radio
structure suggest that the IC process
could be a likely mechanism to generate
most of the extended X--ray flux.

However, a considerable fraction ($\sim 50-60$\%) of the extended emission
flux is confined to the eight--shaped structure (component C in Fig.5)
whose brightness distribution, elongated in a direction at large angle
with the radio--axis and falling down radially roughly as $1/r$, looks very
much different from that expected by a simple IC with the CMB photons.
As a consequence we have first checked whether this feature could be due
to a non spherically symmetric cooling flow.
We find that the addition of a cooling flow to an absorbed
power law source may give an adequate representation of the combined
ROSAT--ASCA spectrum only if a metallicity $\leq 0.2$ solar is assumed.
Acceptable fits, but worse than those of the partial covering
model (Sect.2),
have been obtained with an absorbed 
power law plus constant or non--constant
(s=2 , Mushotzky \& Szymkowiak 1988)
pressure cooling flow models. 
We derive intracluster gas densities between
$0.5-1.3\cdot 10^{-2}cm^{-3}$ 
and, by applying a King's 
model, cluster luminosities $\sim 2-8\cdot 10^{44}$erg s$^{-1}$ in the
0.1--2.4 keV band, that is a factor 15--60 larger than the upper limit
to any thermal contribution within the 2\arcmin ROSAT PSPC
extracted radius (Sect.2).
Furthermore, the cluster emission would give out X--rays in excess of the
observed HRI background by a factor of 
$\sim$1.5--6 at $\sim$1\arcmin distance from the nucleus.  
We conclude that a cooling flow cannot be responsible for the C--component.

We propose that the main features of this component can be
explained by the IC scattering of the IR--optical radiation from a hidden
quasar with the surrounding relativistic electrons of the radio source,
according to the model developed by Brunetti et al.(1997).

\subsection{An IC model of component C}

\par

The IR--optical emission of the nuclear source, as seen by the relativistic
electrons in the radio lobes, is made of two components:
the direct radiation from the quasar, for those electrons located
within the emission cones (assumed 
half--opening angle $\sim 45^{\circ}$), plus
the reprocessed radiation from the dusty molecular torus surrounding
the quasar.
In the model we have fixed 
the inclination of the radio--axis with respect to the sky plane
at $\Psi_{ax}=30^{\circ}$, close to that inferred from
the radio--jet data (Bridle et al.1986), while
the direction of the torus axis ($\theta_T,\phi_T$) is a free parameter 
constrained
by the requirement that the nuclear source is not directly seen 
by the observer ($N_H\sim 2\cdot 10^{21}$cm$^{-2}$) and that
the IC brightness distribution closely matches the 
observations (Fig.6).

\begin{figure}
\centerline{
\psfig{figure=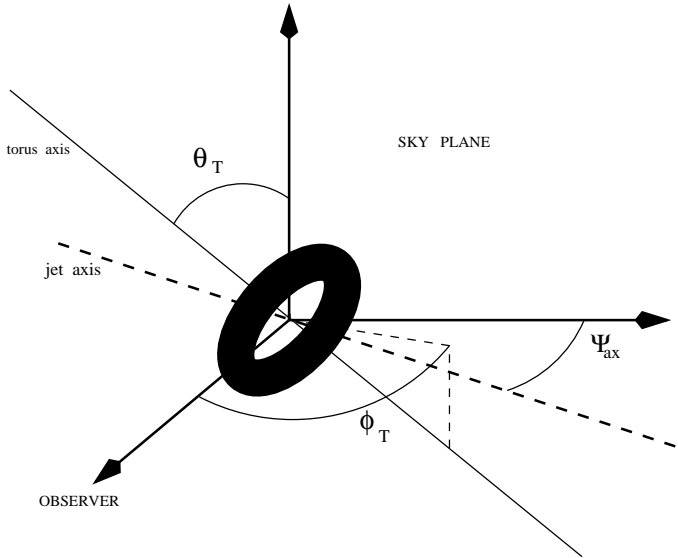,width=9cm,angle=270}
}
\caption{
The model's geometry.
The direction of the torus axis 
in spherical coordinates is given by
($\theta_T$, $\phi_T$); the jet axis is assumed to lie on the 
$\theta_{T}=90^{\circ}$ plane.
The angle between the line of sight and the torus axis
is given by $cos^{-1} (sin \theta_T \, cos \phi_T)$ 
that in the model should be $> 45^{\circ}$.}
\end{figure}

As already stated in the Introduction, a relevant parameter for the 
efficiency of the IC model is the IR radiation from the putative hidden 
quasar.
Unfortunately 3C 219 was not detected in a pointed IRAS observation by Impey
\& Gregorini (1993) and it has not been observed with ISO.
The IR spectral properties of the BLRGs 
observed by IRAS are poorly known.
Heckmann et al. (1994) found that the weighted IRAS 10--86$\mu m$ rest
frame spectral index of a sample of 9 BLRGs given by 
the SUPERSCAMPI procedure
is $\alpha=0.19\pm 0.07$.
However, the few detected BLRGs show a large dispersion in 
the 25--60$\mu m$ spectral indices ranging from
-0.47 to +0.44 (Impey \& Gregorini 1993, Heckmann et al. 1994, 
Golombek et al. 1988).
By assuming an IR spectral shape consistent with the IRAS upper limits
and with the range of observed BLRG spectra and, also, with the predicted
spectra of dusty tori models (Pier \& Krolik 1992),
the 6--100$\mu m$ isotropic IR luminosity
of 3C 219 could be as large as $7.5\cdot 10^{44}$ erg s$^{-1}$.

To estimate the IR luminosity of the hidden quasar we first notice
that the monochromatic $50\mu m$ luminosities
of a sample of low redshift ($0.3<z<0.85$)
radio galaxies are on the average
$\sim 5$ times smaller than those of a sample of quasars
in the same redshift interval (Heckmann et al. 1992).
By applying 
Pier \& Krolik (1992) models to the adopted geometrical
configuration of 3C 219 we
predict an IR luminosity a factor  
3--7 lower than seen from a face--on quasar, 
depending on the torus parameters.
If these ratios and the upper limits for the
observed IR luminosity  of 3C 219 are adopted,
the 6--100$\mu m$ rest frame upper limit on
the luminosity of the hidden quasar would be  
$\sim 5 \cdot 10^{45}$erg s$^{-1}$.

We have also tried to estimate the luminosity of the hidden quasar by
following a different approach.

If a mean optical--X-ray spectral index $\alpha_{ox}\simeq 1.4$  
($\alpha_{ox}\equiv -log(L_{2keV}/L_{250\mu m})/2.605$
; Brunner et al. 1994)
is adopted, then from the 0.1--2.4 keV rest frame luminosity
of $3.6\cdot 10^{44}$ erg s$^{-1}$ ($\Gamma =1.8$)
of the unabsorbed nuclear source, and a typical optical spectral
index $\alpha \simeq 0.6$ (Richstone \& Schmidt 1980),
we derive a 350--650$nm$ rest frame luminosity
$\simeq 8.4 \cdot 10^{44}$ erg s$^{-1}$.
A similar value ($\simeq 9.8\cdot 10^{44}$erg s$^{-1}$)
is obtained by making use of the
correlations between the
radio, optical and X--ray powers of radio loud quasars
(Browne \& Murphy 1987, Kembhavi 1993).

Since  
Heckmann et al. (1992,94) have shown that the
6--100$\mu m$ rest frame luminosity
of radio loud quasars is a factor 6--8 larger than the
350--650$nm$ rest frame optical luminosity,
we find an expected
6--100$\mu m$ luminosity $5-6.7\cdot 10^{45}$erg s$^{-1}$,
larger but close to the upper limit derived in the
first estimate previously discussed.

As a consequence in the IC model described below we will adopt
a 6--100$\mu m$ luminosity $=5\cdot 10^{45}$erg s$^{-1}$ 
for the 3C 219 hidden quasar, which 
corresponds to
an IR--optical 100--0.35$\mu m$ rest frame luminosity of 
$8.8\cdot 10^{45}$erg s$^{-1}$ 
if typical IR--optical spectral parameters are
assumed (Brunetti et al. 1997).

In our model computations, we assume an
emission pattern of the reprocessed IR radiation close 
to that predicted by the theoretical models of Pier \& Krolik (1992).
It should be stressed, however, that, due to the smoothing made by the
PSF of the ROSAT HRI, a precise knowledge of this pattern is
not crucial.
A possible additional beamed IR emission (Hes et al. 1995) has not 
been taken into account,
but being it directional anyway it would affect the calculation only
in a small fraction of the radio--volume.

In order to get some insight into the spatial distribution of the relativistic
particles, we have constructed a 3D contour of the radio--volume 
by making use of the weakest isophote in the 1.4GHz--VLA map of
Clarke et al. (1992). 
The radio galaxy volume is assumed to be symmetric 
around the line joining the mid--points of the weakest radio isophote at
each fixed distance from the nucleus.
We have
deprojected the structure with the inclination angle $\Psi_{ax}=30^{\circ}$
and obtained a virtual 3D model 
of the radio galaxy.
Obviously, this is a rather crude approximation as might be indicated
by the complex structure of the radio isophotes.
Given the spectral properties of the
hidden quasar and those of the relativistic electrons, 
a numerical code computes the
total IC soft X--ray luminosity and brightness distribution projected on the
plane of the sky.
In order to obtain a map to be compared 
with that derived from the observations,
the contribution from the IC scattering of the
CMB photons and the observed background level were added
to the theoretical matrix,
the brightness distribution was convolved with the ROSAT HRI PSF
and smoothed with the same gaussian function used for the data
(Sect.3).

As a first approximation we consider
an uniform distribution of the relativistic
particles and an electron spectrum extended to lower energies with the slope
indicated by the radio spectral index, not modified by radiative and
adiabatic losses.

One can distinguish two regions in the
X--ray brightness distribution of the model (Fig.7):
at small distances from the nucleus (comparable with the minor axis of the
radio galaxy) the X--ray emission distribution 
depends mainly on the nuclear radiation field,
the X--ray axis being that of the quasar illumination cone;
at larger distances from the nucleus the X--ray distribution is mainly
determined by the distribution of the relativistic particles and 
the X--ray emission tends to be more and more aligned with the radio
structure.
The IC X--ray flux from the far lobe can be
considerably larger than
that from the near one depending on the inclination of the radio--axis
on the sky plane.
As shown by Brunetti et al.(1997) in the case of ellipsoidal
radio galaxies,
the predicted ratio for an inclination $\Psi_{ax}=30^{\circ}$
can range up to $\sim 5$, depending on the relative contribution
due to the IC scattering of the CMB photons and on the luminosity of 
the hidden quasar.
This may explain the larger extension of the observed brightness 
distribution toward the northern lobe.
At larger distances from the nucleus, where the nuclear photon
energy density becomes lower than that of the CMB, the IC with the CMB
dominates the X--ray flux and the resulting X--ray brightness distribution
becomes aligned with the radio--axis.

The model reproduces fairly well the observed brightness in the 
extended inner regions where the IC emission is dominated by the scattering
of the radiation from the nuclear source (Fig.5 and Fig.7).

\begin{figure}
\centerline{
\psfig{figure=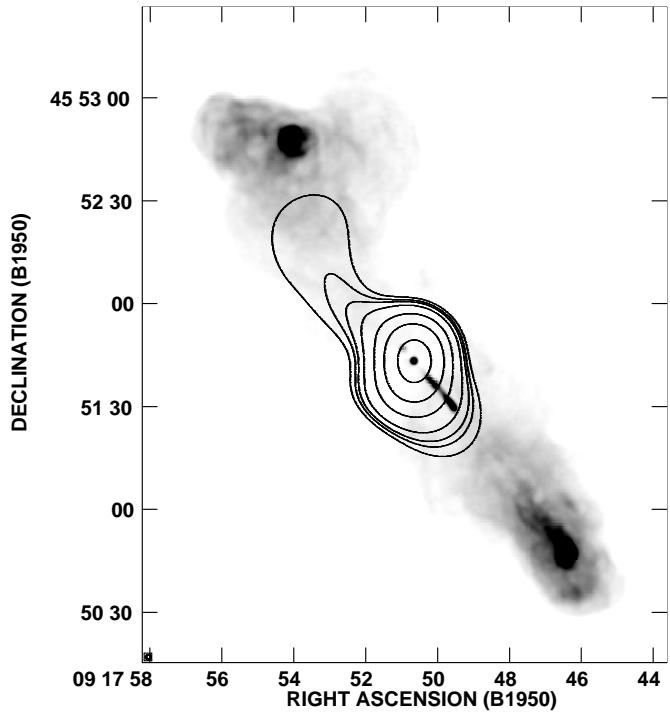,width=9cm}
}
\caption{
The model predicted X--ray isophotes (contours) are
superposed on the VLA 1.4 GHz map (gray--scale).
The inclination of the radio--jet axis on the
sky plane is 30$^{\circ}$,  while for the torus axis 
$\phi_{T} = 69^{\circ}$ and $\theta_T=54^{\circ}$.
%
%
%
%
The isophotes are on the same scale as those
of Fig.5, starting from the 0.24 level, and reproduce
fairly well the C--component.}
\end{figure}

The required inclination of the dusty torus is such that its axis 
makes an angle of $\sim 37^{\circ}$ with the 
radio--axis and $\sim 73^{\circ}$ with the
line of sight.
Large tilts between the radio and dusty lane/torus axis 
for a number of radio galaxies have been 
discovered with ground based telescopes (M\"ollenhoff et al. 1992)
and more recently with HST observations 
(Ford et al. 1994, de Juan et al. 1996).
It also follows that the distribution of the relativistic particles within
the C--component is uniform, at least on the smoothing scale ($\sim 10$ Kpc).

The required relativistic electron density can be compared with that
derived by the minimum energy (equipartition) argument.
Since  relativistic electrons with energies much
lower ($\gamma\leq 1000$) than those of the synchrotron (radio) electrons
dominate our IC model, we apply the equipartition equations given by
Brunetti et al. (1997) for a fixed low energy cut off in the 
electron spectrum (in this paper $\gamma_{min}=50$ and
$B_{eq}\propto \gamma_{min}^{-0.2}$).
The equipartition magnetic field strength evaluated by assuming the
minimum energy 
condition over all the radio--volume 
and equal energy density between negatively and positively charged
particles is $B_{eq}=1\cdot 10^{-5}$G
(with standard equipartition formulae it would be $7\cdot 10^{-6}$G).

Our model requires a density of relativistic particles such that
$B=3 \cdot 10^{-6}$G, i.e. 3.3 times smaller than
the equipartition value.
The minimum energy hypothesis 
is not fulfilled,
the energy in the particles being a factor $\sim$10 larger than
in the equipartition case.
In the case of Fornax A, Feigelson et al.(1995) 
also found a similar, although smaller, 
departure from the equipartition condition, while
in the case of the radio galaxy
PKS 1343--601 (cen B), Tashiro et al.(1998) find
that the energy density
of the relativistic particles ($10^3< \gamma <10^5$),
equally distributed between negative and positive charges,
is a factor $\sim 9$ larger than that 
of the magnetic field.
With the parameters given in Tashiro et al. paper,
by including the energy contribution of mildly relativistic
particles ($\gamma_{min}= 50$), we derive
a ratio $\sim 50$ between particle and magnetic field 
energy densities,  
that is a value of the same order as that found in the
case of 3C 219.

\subsection{The external components}

\par

The simple model discussed so far cannot explain the rather complex
structure seen in the external regions where the IC scattering of the quasar's
photons becomes more and more negligible.
The IC scattering of the CMB photons may explain the observed features under
the assumption that there are deviations from the 
assumed uniformity and simple 3D geometry 
of the spatial distribution of the relativistic particles.
In general, variations of a factor 2--2.5 in the relativistic electron column
densities in excess of those of our model would be sufficient
to give the observed X--ray brightness distribution.
There is some evidence that this might be the case.

Let us consider first the N--component.
Both the radio brightness (Fig.5) and 
the 1.4--5 GHz spectral index distributions (Clarke et al. 1992)
indicate the possible presence of a flow of relativistic
particles toward the E--W direction
from the northern hot--spot to the N--component,
which coincides with a region of steeper radio spectral index.

Since at the position of the N--component our basic IC model predicts
a number of counts $\sim$ 3/4 that of the background (of which $\sim$
20\% from the IC scattering of the nuclear photons), the observed signal can
be generated by an increase of a factor $\sim$2.5 in the number
of relativistic electrons.
We have tested this hypothesis under the assumption that the back--flow
might provide the required enhancement in the density of relativistic
particles leaving the magnetic field strength unchanged, i.e.
a factor $\sim$3.3 lower than the equipartition value of Sect.4.1.
By assuming an electron injection 
spectrum $\delta=2.6$ and radiative
losses, we find a synchrotron radio brightness
of the western part of the north lobe and a 1.4--5 GHz spectral
index $\alpha =1.6-1.7$ both consistent with Clarke et al.(1992)
findings. 
The implied age of the particles reservoir would be
$\sim 4\cdot 10^7$ years.
(For these calculations we have used the SYNAGE package of
Murgia \& Fanti, 1996).

Let us consider now the S--component.
Our basic IC model predicts a number of counts, of which $\sim$35\% from the
IC scattering of the nuclear photons, close to that of the background.
Here an enhancement of a factor of 
2 in the density of relativistic electrons
would be sufficient to account for the observed X--ray flux.

We notice that the S--component lies in a region between the radio
jet and the southern hot--spot where there is evidence of a systematic
steepening of the radio spectral index
and where the magnetic field lines
surround a lobe of brighter radio emission, 
being perpendicular to the line joining the radio--jet with
the hot--spot itself (see the polarization map in Clarke et al. 1992).
This may suggest a strong interaction between two 
relativistic plasmas, one
of which a back--flow from the hot--spot.

We have also considered the possibility that the S and 
N--components are of thermal origin.
The contribution of these components to the total X--ray flux is
$\sim$15\% and we know from the spectral analysis that at most
$\sim$10\% of
the total flux can be thermal ($kT\sim 1.5$ keV).
Thus it is possible that at least one of these components
is of thermal origin.
From the rotation measure (RM) and depolarization maps Clarke et al.(1992)
find that an external clumpy medium is responsible for the observed
RM and depolarization features.
For the sake of clarity we present in Fig.8 the X--ray brightness 
overlayed onto the depolarization map.

\begin{figure}
\centerline{
\psfig{figure=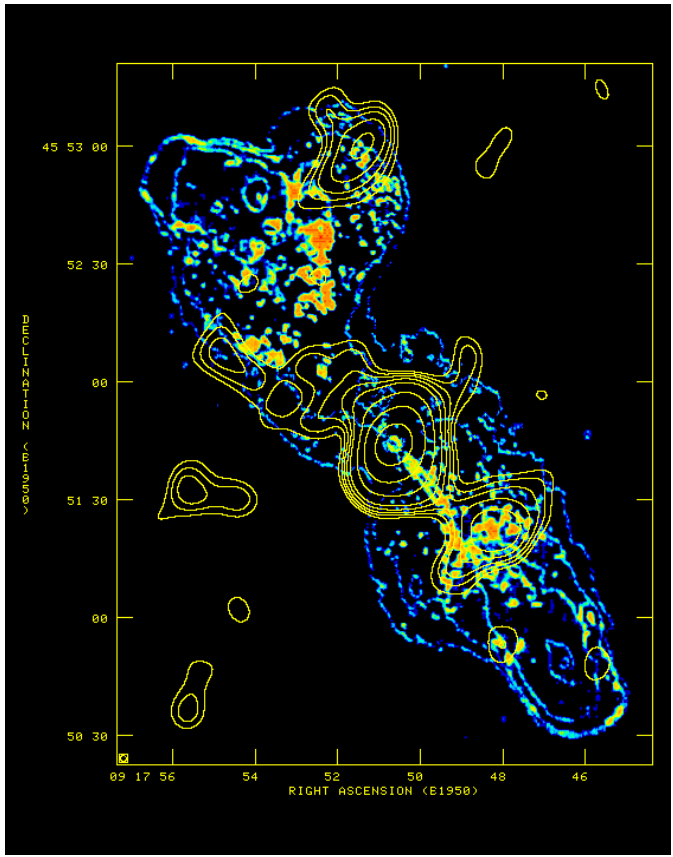,width=9cm}
}
\caption{
The X-ray image of Fig.5 (yellow contours, same levels)
superposed on the high frequency (6 and 18 cm)
depolarization map (color--scale) of 3C 219.
This has been obtained from the  
plate of Clarke et al.(1992) by digitizing  
and transforming in FITS format.
The scale level is in arbitrary units with the most depolarized
regions in red.
The blue contours represent the 22 cm total intensity 
(see Clarke et al.1992).
The coincidence between a region of high depolarization and the S--component
is striking.}
\end{figure}

The coincidence of the S--component with the depolarization structure
transverse to the radio--axis is striking, while a spotted
distribution of moderate depolarization is also observed on the southern
part of the N--component.
On the other hand, we notice that the X--ray isophotes of the
northern radio lobe appear to carefully avoid regions
of larger depolarization.
This is a somewhat contradictory result.
Therefore, we have concentrated our attention on the S--component only.

Let us suppose that,
according to the spectral analysis of Sect.2,
the source is surrounded by a magnetized thermal
plasma with a temperature $\simeq$1.5 keV
and in pressure equilibrium with the
relativistic plasma.
In our non--equipartition model the pressure inside the radio lobes is
$1.4\cdot 10^{-11}$dyne cm$^{-2}$ and the resulting external gas density 
would be $n_e \sim 3 \cdot 10^{-3}$cm$^{-3}$.
The strength of the S--component entails an emission measure
$\sim 3$ pc cm$^{-6}$, that is a depth of the emission region
$\geq 300$ kpc, much larger than the observed feature.
The thermal model may be eased by assuming that the gas is highly
clumped.
We find that a structure with an overall size $\simeq$50 kpc, a filling
factor of a few percentage points and clumps with a mean size $\simeq$ 1 kpc
may account for the observed X--ray intensity and depolarization 
($\sim 0.7$).
A detailed model would require the knowledge of the RM structure
function (Tribble, 1991).
We tentatively conclude that a thermal origin of the S--component cannot
be ruled out.

Finally, we notice 
that the hot--spots are not detected at the sensitivity level of
our HRI observation.
While the binning and smoothing procedures used to enhance the
HRI image statistics give a convolved PSF much 
larger than the southern hot--spot dimension, thus depressing any 
upper limit on the X--ray flux from the hot--spot itself,
the convolved PSF is comparable with the northern hot--spot extension.
The northern hot--spot has been resolved at 1.4 GHz with an angular size
of 12 arcsec (Clarke et al. 1992).
From the analysis of the digitized 1.4 GHz map of 3C 219 we derive a radio flux 
$\simeq$250 mJy contributed by the hot--spot within a circular region
of 90 arcsec$^2$, comparable
with our X--ray beam size, and compute an equipartition ($\gamma_{min}=50$)
magnetic field strength $\simeq 2.7\cdot 10^{-5}$G.
The 2--$\sigma$ upper limit of the X--ray brightness of the hot--spot is 
$\simeq 3.8\cdot 10^{-17}$ erg s$^{-1}$ cm$^{-2}$
arcsec$^{-2}$.
This allows us to set an upper limit
to the density of the relativistic electrons Compton scattering the CMB
photons and a lower bound to the magnetic field of $5.6\cdot 10^{-6}$G,
which is a factor of 4.8 lower than the equipartition value.
Therefore, a deviation from the equipartition condition in the hot--spot 
as well as in the lobes cannot be ruled out.

\section{Conclusions}

\par

A deep ROSAT HRI
X--ray image of the FRII radio galaxy 3C 219 shows
a point--like source, coincident with the nucleus of the galaxy and
accounting for $<$74 \% of the total net counts
($\sim$60 \% of the total net counts by taking into account 
the spectral information), and an emission 
aligned with the radio structure that extends from the 
innermost region of the radio galaxy to the hundred--kpc scale.

The 0.1--10 keV spectrum is well represented by a partial
covering model in which $\sim$ 74\% of a 
power law spectrum ($\Gamma=1.8$) is absorbed by a column density 
$N_H \simeq 2-3 \cdot 10^{21}$ cm$^{-2}$.
This spectral slope coincides with that of the synchrotron emission of the 
radio lobes.
The spectral analysis indicates that at most 10\% of the X--ray flux 
could be due to a thermalized gas with a temperature of
$\sim$1.5 keV.
The point--like source, whose de--absorbed (isotropic) luminosity
in the 0.1--2.4 keV band is $3.6\cdot 10^{44}$erg s$^{-1}$, can be
identified with the emission of a quasar hidden in the nucleus of 
3C 219.
This lends further support to the unification of FRII radio galaxies
with radio loud quasars.

Subtraction of the point--like source leaves an extended, unabsorbed 
circumnuclear component (named C) whose luminosity in the 0.1--2.4 keV
band is $\simeq 2 \cdot 10^{43}$erg s$^{-1}$.
It appears unlikely that a cooling flow may explain this component
because the X--ray emission from the associated intracluster gas
would exceed by far the observed counts.
As a matter of fact, we have no evidence of a diffuse hot
intracluster gas.

In agreement with the unification scheme, we have constructed a simplified
model in which the main features of 
the C--component  can be satisfactorily
explained as IC scattering of the IR--optical radiation from the
hidden quasar, and surrounding dusty/molecular torus, with the relativistic
electrons uniformly distributed in the radio lobes.
Since the
C--component  is significantly inclined with respect to the direction of
the radio--jet, our model implies a large tilt ($\sim 40^{\circ}$) 
between the radio 
jet and the torus axis.
Because of the limits one can place on the luminosity of the hidden quasar,
we find that the X--ray flux of the C--component can be accounted for if the
equipartition condition is violated, the energy in the relativistic particles
being at least a factor 10 larger than that in equipartition conditions.
The electrons mainly involved in the IC production of the soft X--rays
have energies much lower ($\gamma \sim 100-300$) than those of the electrons
producing the synchrotron radio emission.
Therefore, if our model is correct, it can provide useful constraints
on the particle acceleration mechanisms and aging of the radio source.

The X--ray structures (named S and N--components) observed in the outer
regions cannot be explained by our model in which the relativistic 
particles are uniformly distributed within the radio lobes
(it should be borne in mind that we have consistently included in our
computations the contribution from the IC scattering of the CMB photons,
which is of minor importance for the C--component, but dominant
as one goes further out in the radio lobes).
Accounting for these structures by the IC process requires positive 
fluctuations of a factor 2--2.5 in the column densities of the relativistic
electrons.
We tentatively associate this possibility with the evidence of
back flows from the hot--spots.
Alternatively, we cannot exclude the possibility of a thermal contribution
from a hot (kT $\sim$ 1.5 keV) clumpy gas surrounding the radio lobes
in localized regions.
The presence of
moderate depolarization and RM gradients from the radio maps
supports this hypothesis
in the case of the S--component.
Observations with AXAF will provide an invaluable tool to
verify our model and the nature of the X--ray emission of 3C 219.

\begin{acknowledgements}

We warmly thank Ingo Lehmann and G\"unther Hasinger for their help
in the wobble correction procedure for the analysis of the HRI data, 
and Gianni Zamorani for helpful discussions.
We also thanks the referee for halpful comments.
The radio image has been retrieved by the web archive of the DRAGNs 
by Leahy, Bridle and Strom (http://www.jb.man.ac.uk/atlas/).   
This research has made use of the NASA/IPAC Extragalactic Database
(NED) which is operating by the Jet Propulsion Laboratory, California
Institute of Technology under contract with the National Areonautic
and Space Administration.
This work has been partially supported by the Italian Space Agency
(ASI--ARS--96--70).
\end{acknowledgements}

\end{document}